\documentclass[11pt]{article}

\usepackage[english]{babel}
\usepackage[a4paper,nohead,twoside]{geometry}
\usepackage{amsmath}
\usepackage[mathscr]{euscript}
\usepackage{amsfonts}
\usepackage{amssymb}
\usepackage{amsthm}
\usepackage{mathtools}
\usepackage{graphicx}
\usepackage{color}
\usepackage{enumitem}
\usepackage{xspace}

\usepackage[bookmarks=true,
            bookmarksopen=true,colorlinks=false,breaklinks=true]{hyperref}

\newtheorem{theorem}{Theorem}[section]

\newtheorem{conjecture}[theorem]{Conjecture}
\numberwithin{equation}{section}

\newcommand \del 		\partial

\begin{document}

\title{On Strong Cosmic Censorship}
\author{James Isenberg\footnote{Department of Mathematics and Institute of Theoretical Science, University of Oregon. \newline  isenberg@uoregon.edu}}

\date{\today}
\maketitle

\begin{abstract}
For  almost half of the one hundred year history of Einstein's theory of general relativity, Strong Cosmic Censorship has been one of its most intriguing conjectures. The SCC conjecture addresses the issue of the nature of the singularities found in most solutions of Einstein's gravitational field equations: Are such singularities generically characterized by unbounded curvature? Is the existence of a Cauchy horizon (and the accompanying extensions into spacetime regions in which determinism fails) an unstable feature of solutions of Einstein's equations? In this short review article, after briefly commenting on the history of the SCC conjecture, we survey some of the progress made in research directed either toward supporting SCC or toward uncovering some of its weaknesses. We focus in particular on model versions of SCC which have been proven for restricted families of spacetimes (e.g., the Gowdy spacetimes), and the role played by the generic presence of Asymptotically Velocity Term Dominated behavior in these solutions. We also note recent work on spacetimes containing weak null singularities, and their relevance for the SCC conjecture.

\end{abstract}

%Outline of a paper aimed for CQG.

% TABLE OF CONTENTS
\tableofcontents

% \newpage

% % LIST OF TASKS
% \listoftasks

\newpage

% START OF PAPER
%
\section{Introduction}
\label{Intro}
    Ever since the 1916 discovery of the Schwarzschild solution, singularities have played a major role in general relativity. During the first fifty years following Einstein's proposal of general relativity in 1915, singularities appeared primarily as a mathematical feature of most of the explicit solutions of Einstein's gravitational field equations found in that period: In the Friedmann-Lema\^{\i}tre-Robertson-Walker (FLRW), the Kasner, the Taub, and the Kerr solutions, as well as that of Schwarzschild, it was found that if the coordinates are extended to their natural limits, the metrics of these solutions either blow up or go to zero, thereby ``going singular". Such behavior was familiar to physicists in solutions of the more familiar theories such as Maxwell's theory of electromagnetism\footnote{Physicists in the 1800s worried  about the singularity at the origin of the Coulomb solution of Maxwell's equations. These worries were quieted by the recognition of the quantum nature of ``point sources" for electromagnetic fields, such as the electron.}. However in Maxwell's theory, many solutions without singularities were well-known as well. With singular solutions appearing to dominate in general relativity, many asked if in fact they are a generic feature of the theory, or rather just an artifact of the symmetries of the known solutions. 

The issue of the genericity of singularities in solutions of Einstein's equations was a major topic of research in general relativity during the 1960s. A number of approaches to explore this issue were pursued; the one which (in a certain sense) led to a definitive answer, is the one based on the dynamics  of congruences of causal geodesic paths. Using this approach, Hawking and Penrose proved a number of results (see Section \ref{HPSing}) which, suitably interpreted, claim that singularities occur in generic solutions. These results, which are
often collectively labeled as the ``Hawking-Penrose Singularity Theorems", lead to this claim \emph{if} one interprets ``singularity" to be causal geodesic incompleteness (CGI), and \emph{if} one interprets the hypotheses of the Hawking-Penrose theorems as corresponding to a generic class of solutions. 

While causal geodesic incompleteness indicates  some degree of pathology in a spacetime with this characteristic, the features accompanying CGI vary widely from spacetime to spacetime. For example, in the FLRW and the Kasner solutions, curvature and tidal force blowups accompany the CGI; while in the Taub-NUT extension of the Taub spacetime, the curvature is bounded, but the presence of a Cauchy horizon is closely tied to causal geodesic incompleteness. In terms of what an observer sees along his or her worldline, those proceeding toward the singularity in an FLRW or Kasner spacetime are stretched and crushed to death, while those heading for the singularity in a Taub-NUT spacetime may enter a region in which the ability to predict the future from a known set of initial conditions breaks down. 

Whether or not one or the other of these behaviors is ``physically preferable" or in some sense ``less singular", it would be very interesting to determine which happens more often. Penrose addressed this question almost fifty years ago, proposing his Strong Cosmic Censorship (SCC) conjecture. Roughly stated (as was the case in its early form), the SCC conjecture claims that in solutions of Einstein's equations, curvature blowup generically accompanies causal geodesic incompleteness. 

Both because of its mathematical elusiveness and because of its somewhat compelling physical implications (should we expect to be crushed, or might we be able to ``go back in time"?), the Strong Cosmic Censorship conjecture has been viewed as one of the central questions of mathematical relativity.  Despite this strong interest, SCC remains unresolved, and one might argue that very little of the work which has been done to date tells us anything directly regarding whether SCC holds or not. Much of this work has been devoted to the study (and proof) of model versions of Strong Cosmic Censorship, which are either restricted to families of solutions characterized by each spacetime in the family admitting a nontrivial isometry group of dimension one or higher (e.g., solutions which are spatially homogeneous, or are invariant under a spatially acting $T^2$ or $U(1)$ isometry group), or are restricted to sets of solutions which are small perturbations of known solutions (e.g., perturbations of an FLRW spacetime \cite{RS}). Since SCC is fundamentally about behavior which is generic in the set of \emph{all} solutions of Einstein's equations, and since these isometry-based families are effectively of measure zero in this set, a model proof of SCC in one of these families is not directly related to the validity of the full SCC conjecture. However, it is hoped that these model studies do allow researchers to develop tools that could be useful in the study of the full conjecture. 

We begin this  report on the Strong Cosmic Censorship conjecture with a brief review (in Section \ref{HPSing}) of what the Hawking and Penrose theorems explicitly tell us regarding the prevalence of singularities---in the sense of geodesic incompleteness---in solutions of Einstein's equations. Next, in Section \ref{Penrose}, we discuss Penrose's original conception of Strong Cosmic Censorship. In doing this, we also comment on his conception of Weak Cosmic Censorship (WCC), noting that SCC does not imply WCC, and WCC does not imply SCC. 

In these earliest formulations, Penrose did not seek to state either conjecture in a rigorous way. We provide an example of a rigorous statement of SCC  in Section \ref{PoldGowdy}, where we discuss a model Strong Cosmic Censorship theorem for the limited case of Polarized Gowdy spacetimes. This  discussion introduces the idea of studying SCC by stating and proving model SCC-type theorems for limited families of spacetimes. We continue in this direction in Section \ref{TGowdy}, where we discuss some of the ideas used in Ringstr\"{o}m's significantly deeper and more intricate proof of a model SCC theorem for all ($T^3$) Gowdy spacetimes.

%survey the collection of families of spacetimes for which an SCC-type theorem has been proven to date, highlighting  Ringstrom's SCC theorem for all ($T^3$) Gowdy spacetimes. 

One of the key steps in the proofs of these model SCC theorems, both for the polarized and for the general Gowdy spacetimes, is the demonstration that generic $T^3$ Gowdy spacetimes exhibit \emph{asymptotically velocity-term dominated} (AVTD) behavior. We discuss in Section \ref{PoldGowdy} what AVTD behavior is, and how it can be useful in studying SCC. Then in Section \ref{AVTD}, we discuss evidence for (and against) the presence of AVTD behavior in a number of families of spacetime solutions of Einstein's equations. 

In his early discussions of SCC, one of Penrose's main arguments for the conjecture was that ``blue shift" effects would tend to disrupt the formation of a Cauchy horizon inside black holes. Christodoulou, Dafermos, and others have developed this idea into an approach for studying SCC which has recently yielded considerable insight. A key feature of this approach is the possible development of \emph{weak null singularities} inside black holes. We discuss in Section \ref{Blue} what these are, their possible stability, and the implications for SCC if indeed weak null singularities are a stable feature of solutions of Einstein's equations. We conclude with comments on the fruitful role the Strong Cosmic Censorship has played in mathematical relativity.

\section{The Hawking-Penrose Singularity Theorems}
\label{HPSing}

The notion of a singularity is difficult to pin down in general relativity because, in contrast to Maxwell's theory or Navier-Stokes' theory for which there is an a priori fixed background spacetime on which to check whether or not the fields are bounded, for Einstein's theory the spacetime on which the fields are defined is not fixed. Sets on which the fields blow up can be removed or added; hence the presence or absence of unbounded fields in a given solution is a malleable feature. As well, the boundedness of the gravitational field in a given solution can depend strongly on the choice of frame and the choice of coordinates. These difficulties led researchers during the 1960s to settle on causal geodesic incompleteness of an inextendible spacetime\footnote{Here $M^{n+1}$ is a spacetime manifold, $g$ is a Lorentz-signature spacetime metric, and $\Psi$ collectively represents the non-gravitational fields} $(M^{n+1}, g, \Psi)$ to be the criterion for labeling that spacetime as singular. (See \cite{Ger} for further discussion and justification of this criterion.) 

If one decides to use  CGI as the mark of a singular solution of the Einstein equations, the study of congruences of geodesic paths in them is a natural way to determine if singularities are a prevalent feature of such solutions. One of the key tools for studying geodesic congruences  is the Raychaudhuri equation, which (for a surface-orthogonal timelike congruence) takes the form 
\begin{equation}
\label{Raychaud}
\nabla_U \Theta = - \frac 14 \Theta^2 - \Sigma_{\alpha \beta} \Sigma ^{\alpha \beta}-R_{\alpha \beta}U^{\alpha} U^{\beta}; 
\end{equation} 
here $U$ is the vector field tangent to the congruence, $\nabla_U$ is the directional derivative along $U$, $\Theta:=\nabla_\alpha U^\alpha $ is the expansion of the congruence, $\Sigma_{\alpha \beta} := \frac 12 (\nabla_\alpha U_\beta + \nabla_\beta U_\alpha)- \frac 14 \Theta g_{\alpha \beta} $ is the shear of the congruence, and $R_{\alpha \beta}$ is the Ricci curvature tensor of the spacetime.

As stated here, this equation has nothing to do with whether or not the spacetime containing the geodesic congruence is a solution of Einstein's equations; it is purely a geometric consequence of tracing over the definition of the Riemann curvature tensor. However, if one uses the Einstein field equations in the form
\begin{equation}
\label{RicEinsteqs}
R_{\alpha \beta} = \kappa (T_{\alpha \beta} - \frac 12 g_{\alpha \beta} T) + \Lambda g_{\alpha \beta} 
\end{equation} 
to replace the Ricci tensor in the Raychaudhuri equation \eqref{Raychaud}, and if one presumes that the stress-energy tensor $T_{\alpha \beta}$ (together with the cosmological constant $\Lambda$, and Newton's constant $\kappa$) satisfies the positivity condition\footnote{This positivity condition has been labeled the ``strong energy condition".} 
\begin{equation}
\label{strongenergycond}
\kappa(T_{\alpha \beta} W^\alpha W^{\beta} - T^\beta_\beta W_\alpha W^\alpha ) + \Lambda W_\alpha W^\alpha>0
\end{equation}
for any timelike vector field $W$, then the Raychaudhuri equation tells us that if the congruence expansion $\Theta$ is non-zero at any point $p$ for some congruence of surface-orthogonal paths, then $\Theta$ must blow up in finite proper (affine) time either to the future (if $\Theta(p)>0$) or to the past (if $\Theta(p)<0$) of $p$. This result is one of the primary tools used for proving most of the Hawking-Penrose singularity theorems.

A wide variety of different results proven during the 1960s  are collectively known as the Hawking-Penrose singularity theorems. Just about all of them have a particular characteristic form: They show that a spacetime $(M^{n+1}, g, \Psi)$ must be causal geodesically incomplete (and hence ``singular") so long as that spacetime satisfies a set of conditions including each of the following: (i) a causality condition (e.g., the spacetime admits no closed causal paths); (ii) a regularity condition (e.g., the spacetime is smooth); (iii) an ``energy condition" (e.g., the spacetime satisfies \eqref{strongenergycond}); (iv) a curvature ``generic condition" (e.g., every causal geodesic in the spacetime contains at least one point at which 
\begin{equation}
\label{generic}
V_{[\gamma} R_{\alpha] \mu \nu [\beta}V_{\delta] }V^\mu V^\nu \neq 0,
\end{equation}
with  $V^{\mu}$ the vector tangent to the geodesic, and with $V_{[\alpha} W_{\beta]}$ indicating index skew-symmetrization); and (v) a boundary/initial condition (e.g., the spacetime admits a closed achronal hypersurface).

The following archetypal example of such a theorem, proven by Hawking and Penrose in 1970 \cite{HP70} follows this pattern closely:
\begin{theorem}[Hawking-Penrose Singularity Theorem]
\label{HPThm}
If a spacetime $(M^{3+1},g,\Psi)$ with stress-energy $T^{\alpha \beta}$ is a smooth solution of Einstein's equations, if it contains no closed timelike paths, if it satisfies the strong energy condition \eqref{strongenergycond}, if the inequality \eqref{generic} holds at least somewhere along every one of its causal geodesic paths, and if it admits either a closed achronal hypersurface or a closed trapped surface, then the spacetime cannot be causal geodesically complete.
\end{theorem}

Does this, or any other such theorem, show that ``generic" solutions are CGI, or even that ``physically reasonable" solutions are generically CGI? Of course deciding this one way or the other depends on what one means in using this terminology, and how such meaning compares with the conditions contained in the hypothesis of Theorem \ref{HPThm}. In exploring the behavior of solutions of Einstein's equations, it is not generally considered to be overly restrictive to eliminate those solutions which fail to satisfy a causality condition or fail to be differentiable in some appropriate sense.\footnote{The spacetime should be sufficiently differentiable so that geodesic incompleteness does not arise simply because the spacetime is not smooth enough to admit a geodesic congruence.} The genericity of the other three conditions is less convincing, however: While the strong energy condition \eqref{strongenergycond} does hold for Einstein-vacuum as well as Einstein-Maxwell solutions, it fails for solutions with negative cosmological constant (the sign which is needed for simple cosmological models with accelerated expansion). One may reasonably choose to focus on spacetimes containing a closed achronal hypersurface or even a closed Cauchy surface; but if one is interested in asymptotically flat solutions, the presumption that there is an embedded trapped surface is somewhat restrictive\footnote{See, however, the work of Christodoulou \cite{Chris} and of Klainerman and Rodnianski \cite{KlainRod} in which conditions on initial data are given which guarantee that a trapped surface will form in the spacetime development of that data.}. 
As for the condition that inequality \eqref{generic} hold somewhere along every causal geodesic path, although this is often labeled ``the generic condition" by those using it, there is no particular evidence one way or the other that this condition is indeed generic.

Whether or not it follows from the singularity theorems  that solutions of Einstein's equation are generically CGI, this issue is not crucial in considering Strong Cosmic Censorship. SCC is concerned with generic behavior in spacetimes which contain incomplete causal geodesics, not whether the CGI property itself is generic.

\section{Penrose's Cosmic Censorship Conjectures}
\label{Penrose}

Soon after proving the first of the Hawking-Penrose singularity theorems, Penrose began discussing ideas which evolved into the cosmic censorship conjectures. The first appearance of these ideas in the literature was in \cite{Pen68} and \cite{Pen69} in the late 1960s. While these references do not present a definitive statement of SCC, they do provide an intuitive formulation:

\begin{conjecture} [Intuitive Version of Strong Cosmic Censorship]
Globally hyperbolic spacetime solutions of Einstein's equations generically cannot be extended as solutions past a Cauchy horizon\footnote{A Cauchy horizon in a spacetime $(M,g)$ is a null hypersurface which divides the spacetime into a region which is globally hyperbolic, and a region which is not.}.
\end{conjecture}

Simultaneous with his discussions of SCC, Penrose proposed a second, very different, but equally intriguing conjecture. Labeled Weak Cosmic Censorship (WCC), this conjecture takes the following intuitive form: 

\begin{conjecture} [Intuitive Version of Weak Cosmic Censorship]
In generic asymptotically flat spacetime solutions of Einstein's equations, singularities are contained within black hole horizons.
\end{conjecture}

Our focus in this review is on SCC, not WCC. We mention the latter here primarily to emphasize the fact that neither conjecture (if proven) implies the other--they are logically distinct. We also note that the shared name ``cosmic censorship" pertains more aptly to WCC  than to SCC: Weak cosmic censorship proposes that ``naked singularities"--those visible to far away observers--do not occur generically (they are forbidden by the ``cosmic censor"). 

One key shared feature of SCC and WCC is that both concern the behavior of \emph{generic} solutions. While this term must be made precise before either conjecture can be proven, even in rough form the implication is clear that the existence of solutions with Cauchy horizons does not disprove SCC, and the existence of asymptotically flat solutions with singularities to the causal past of asymptotic observers does not disprove WCC.  This important feature invalidates the majority of the proffered counterexamples to both WCC and SCC which have appeared in the literature. 

\section{A Model SCC Theorem: Polarized Gowdy Spacetimes}
\label{PoldGowdy}
One way to explore evidence favoring or disfavoring a comprehensive conjecture such as SCC is to study if a suitably adapted form of it is valid for special families of solutions. Presuming that these families are essentially of measure zero in the space of all solutions, such studies can neither prove nor disprove the conjecture.  However, in attempting to prove or disprove model versions of SCC (``model-SCC") in special families (such as the Gowdy spacetimes) one can develop ideas, techniques, and scenarios which might ultimately be useful in determining if in fact the SCC conjecture holds. Of course, one must also keep in mind that the lessons learned in proving model-SCC for a given family could instead be misleading, as we discuss below.

While model versions of SCC have been proven for larger families of spacetimes,  the  family of polarized Gowdy solutions provides a very good example of the rigorous formulation and proof of such a result. Hence, we discuss some of the details of this case here.

The polarized Gowdy spacetimes are solutions of the vacuum Einstein's equations which are characterized by the following geometric features: 1) Each solution admits an effective $T^2$ isometry group acting spatially (hence there are two independent commuting everywhere-spacelike Killing fields). 2) The two Killing fields can be aligned orthogonally everywhere (this is the ``polarizing condition";  without it, one has a general Gowdy spacetime). 3) The Killing field ``twists",   which take the form $X \wedge Y \wedge dX$ and $X \wedge Y \wedge dY$ for $X$ and $Y$ labeling the one-forms corresponding to the Killing fields, vanish. 4) The spacetimes admit compact Cauchy surfaces. 

These conditions allow for three possible spacetime  manifolds: $T^3 \times R^1, S^2\times S^1 \times R^1$, and $S^3 \times R^1$ (Lens spaces may also replace the 3-sphere; however the analysis is no different for such  replacements). While a model-SCC has been proven for all of these cases \cite{CIM}, to avoid unnecessary detail here, we restrict our discussion here to the $T^3 \times R^1$ case. 

For polarized Gowdy spacetimes, coordinates may be chosen so that the metric can be written as follows:
\begin{equation}
\label{PolGowdyMetric}
g=e^{\frac{(\tau+\lambda)}{2}}(-e^{-2\tau}d\tau^2 +d\theta^2) +e^{-\tau} (e^P dx^2 +e^{-P}dy^2). 
\end{equation}
Here $\tau \in R^1$ and $(\theta, x,y)$ are coordinates on the 3-torus, with the orbits of the $T^2$ isometry group corresponding to 2-surfaces of constant $\tau$ and constant $\theta$. In terms of the metric functions $P$ and $\lambda$ (which are functions of  $\theta$ and $\tau$ only), the vacuum Einstein equations take the form 
\begin{align}
\label{Peqn}
\partial_{\tau \tau} P &=  e^{-2\tau}\partial_{\theta \theta} P,\\
\label{lambdatau}
\partial_\tau \lambda &=  (\partial_\tau P)^2 +e^{-2\tau} (\partial_\theta P)^2,\\
\label{lambdatheta}
\partial_\theta \lambda &=  2 \partial_\tau P \partial_\theta P.
\end{align}
It is readily apparent from the form of these equations that the initial value problem for the polarized Gowdy spacetimes is well-posed. In particular, one sees that for any choice of a smooth pair of functions $P(\theta,0) =p(\theta)$ and $\partial_\tau P(\theta,0)=\pi(\theta)$ satisfying the integrability condition $\int_{S^1} \pi \partial_\theta p
d\theta =0$  on the circle, the wave equation \eqref{Peqn} admits a unique (maximally extended) solution $P(\theta, \tau)$; and for any choice of a constant $\lambda(0,0)=\ell$ together with the solution for $P(\theta, \tau)$, equation \eqref{lambdatheta} can be integrated to produce initial data $\lambda(\theta,0)$ for $\lambda$, after which \eqref{lambdatau} can be used to evolve to a unique (maximally extended) solution $\lambda(\theta, \tau)$.  Letting $\Pi_{pol}$ denote the space of initial data sets (a pair of functions on the circle plus a constant, with the functions satisfying the integrability condition noted above) for the polarized Gowdy spacetimes, and noting that the evolution of $(P(\theta, \tau), \lambda(\theta, \tau))$ from a set of initial data as just described corresponds to the unique maximal globally hyperbolic spacetime development \cite{CB-G} of that data set,  we may state a model SCC theorem for these solutions as follows: 

\begin{theorem} [Model-SCC Theorem for Polarized Gowdy Spacetimes]
\label{SCCPoldGowdy}
There exists an open dense subset (in $C^\infty$ topology)  $\hat \Pi_{pol}$ of $\Pi_{pol}$ such that the maximal globally hyperbolic spacetime development of any data set in $\hat \Pi_{pol}$ is inextendible.
\end{theorem}

%We notice that these field equations for the polarized Gowdy spacetimes are semi-decoupled: Once initial values for $P$ and $\partial_\tau P$ have been chosen, equation \eqref{Peqn} may be solved as an evolution equation for $P$, independent of $\lambda$. After solving for $P$, one may use \eqref{lambdatheta} to obtain initial data for $\lambda$, and then use \eqref{lambdatau} to evolve $\lambda$ into the future and the past.

The key to proving this theorem is to first show that the AVTD property holds for all polarized Gowdy spacetimes; then to use this result to show that there is a homeomorphism from $\Pi_{pol}$ to a space $\mathcal Q_{pol}$ consisting of sets of asymptotic data which characterize the behavior of the solution approaching the singularity; and finally to show that for an open and dense subset $\hat {\mathcal Q}_{pol}$ of $\mathcal Q_{pol}$, the curvature blows up in a neighborhood of the singularity, hence preventing extension across a Cauchy horizon in that singular region. In
addition, one must show that in the expanding direction of these $T^3$ polarized Gowdy solutions\footnote{In terms of the areal coordinates for the $T^3$ Gowdy spacetimes, which we use here in writing the metric in the form \eqref{PolGowdyMetric}, the singularity occurs at $\tau \rightarrow \infty$, and the spacetime expands with decreasing $\tau$. The $S^3$ and $S^2\times S^1$ Gowdy spacetimes are singular both to the future and the past.}, the spacetimes are geodesically complete and nonsingular. 

Since AVTD behavior plays such a central role in the study of SCC for these as well as other spacetimes discussed below, it is useful to describe the property and how it is verified for these spacetimes in a bit more detail. To define AVTD behavior for a family of solutions of Einstein's equations, we need to first determine an associated VTD system of equations. In the case of the polarized Gowdy solutions, written in areal coordinates, the associated VTD equations are the same as the full set of Einstein's equations \eqref{Peqn}-\eqref{lambdatheta}, but with the spatial derivatives dropped from the first two equations\footnote{Spatial derivatives are dropped from the first two equations, since the idea is that spatial derivatives are dominated by temporal derivatives. In the third equation, \eqref{lambdatheta}, there are no temporal derivatives, so the spatial derivatives are not neglected.}
\begin{align}
\label{PeqnV}
\partial_{\tau \tau} \tilde P &= 0,\\
\label{lambdatauV}
\partial_\tau \tilde \lambda &=  (\partial_\tau \tilde P)^2 ,\\
\label{lambdathetaV}
\partial_\theta \tilde \lambda &=  2 \partial_\tau \tilde P \partial_\theta \tilde P.
\end{align}
Noting that the singularity for these spacetimes occurs as $\tau$ approaches $ + \infty$, we define a particular polarized Gowdy solution $(P(\theta, \tau), \lambda(\theta, \tau))$ to have AVTD behavior if there exists a solution $(\tilde P(\theta, \tau), \tilde \lambda(\theta, \tau))$ of the VTD equations such that the solution of the full system rapidly approaches the VTD solution for large $\tau$. In fact for the polarized Gowdy solutions, it has been shown \cite{CIM} that there exists a constant $C$ such that $|P(\theta, \tau) - \tilde P(\theta, \tau)|< C e^{-2 \tau}$ and $|\lambda (\theta, \tau) - \tilde \lambda (\theta, \tau)|< C e^{-2 \tau}$.

The polarized Gowdy VTD equations \eqref{PeqnV}-\eqref{lambdathetaV} are simple enough that it is very easy to determine the form of the sets of asymptotic data which comprise the space $\mathcal Q_{pol}$. Since the general solution to \eqref{PeqnV} is $\tilde P(\theta, \tau) = v(\theta) \tau +\phi (\theta)$ for an arbitrary pair of (smooth) functions $v$ and $\phi$ which satisfy the integrability condition $\int_{S^1}v \frac{d \phi}{d\theta}d\theta =0$ on the circle, and since the solution $\tilde \lambda (\theta, \tau)$ is readily obtained by simply integrating \eqref{lambdatauV} and \eqref{lambdathetaV} with a single specified constant, the space $\mathcal Q_{pol}$ of asymptotic data consists of choices of $v(\theta)$ and $\phi(\theta)$ satisfying the integrability condition, plus a constant. 

The proof that all solutions of the polarized Gowdy equations do exhibit AVTD behavior in the sense described above is a relatively straightforward consequence of the verification \cite{IM90} that each of the sequence of energy functionals 
\begin{equation} 
\label{PolGowdyEnergies} 
E_k = \sum_{j \leq k-1} \int_{S^1} [ \frac{1}{2}(\partial^j_\theta \partial_\tau P)^2 + \frac{1}{2}(\partial^{j+1}_\theta P)^2 ] d \theta
\end{equation}
monotonically decays in time $\tau$. This monotonicity, together with Sobolev embedding, allows one to control the growth of $P$ and its derivatives, from which the convergence result and the consequent  verification of AVTD behavior follow. The bijectivity and continuity of the map from $\Pi_{pol}$ to $\mathcal Q_{pol}$ readily follow from these estimates as well.

How do we infer results concerning the generic inextendibility of polarized Gowdy solutions from these AVTD results? As shown in \cite{IM90}, if one writes the general solutions to the polarized Gowdy equations in the form of solutions of the VTD equations plus strongly controlled remainder terms, one can calculate the spacetime curvature polynomial scalars (including the Kretschmann scalar)  in terms of the asymptotic data $(v(\theta), \phi(\theta))$, and determine that these invariants fail to blow up along an observer path approaching the limiting spatial coordinate $\theta_0$ \emph{only} if $v^2(\theta_0)=1$, $\frac{d v}{d\theta}(\theta_0)=0$, and $\frac{d^2v}{d\theta^2}(\theta_0)=0$ all hold. The collection of solutions which cannot be extended past the singularity at $\tau \rightarrow \infty$ because of curvature blowup corresponds to all sets of asymptotic data in $\mathcal Q_{pol}$ which do not satisfy these conditions; clearly this set is open and dense among the set of all solutions. 

In closing this discussion of the verification of a model version of Strong Cosmic Censorship for the polarized Gowdy spacetimes, we note that while the proof of these results relies heavily on the verification that the AVTD property holds for these  spacetimes, and while the AVTD property is defined and verified with respect to a particular choice of coordinates (areal coordinates here), neither the statement of Theorem \ref{SCCPoldGowdy} nor its validity depends on a choice of coordinates.

\section{A Model SCC Theorem: $T^3$ Gowdy Spacetimes}
\label{TGowdy}
The difference between the polarized Gowdy spacetimes and the general Gowdy spacetimes is the presence in the latter of an extra (off-diagonal) metric coefficient in the Killing field orbits; specifically, in its $T^3$ version, the metric takes the form  
\begin{equation}
\label{GowdyMetric}
g=e^{\frac{(\tau+\lambda)}{2}}(-e^{-2\tau}d\tau^2 +d\theta^2) +e^{-\tau} [e^P dx^2 + 2 e^P Q dxdy +(e^{-P} +e^PQ^2) dy^2], 
\end{equation}
and the vacuum Einstein field equations take the form
\begin{align}
\label{GPeqn}
\partial_{\tau \tau} P &=  e^{-2\tau}\partial_{\theta \theta} P + e^{2P}(\partial_\tau Q^2 -e^{-2\tau} \partial_\theta Q^2),\\
\label{Qeqn}
\partial_{\tau \tau}Q &= e^{-2\tau}\partial_{\theta \theta} Q -2(\partial_\tau P \partial_\tau Q- e^{-2\tau} \partial _\theta P \partial _\theta Q),\\
\label{Glambdatau}
\partial_\tau \lambda &=  (\partial_\tau P)^2 +e^{-2\tau} (\partial_\theta P)^2 +e^{2P}(\partial_ \tau Q^2 +e^{-2 \tau} \partial_\theta Q^2) ,\\
\label{Glambdatheta}
\partial_\theta \lambda &=  2 (\partial_\tau P \partial_\theta P + e^{2P} \partial _\tau Q \partial_\theta Q).
\end{align}
We note that the addition of the dynamical variable $Q(\theta, \tau)$ clearly complicates the dynamics of the Gowdy spacetimes, but still leaves the function $\lambda(\theta, \tau)$ in a subsidiary role: One can solve for $P(\theta, \tau)$ and $ Q(\theta, \tau)$ independently of $\lambda$, and then obtain the latter by integrating \eqref{Glambdatheta} and then \eqref{Glambdatau}. We note as well that, as in the polarized Gowdy case, an initial data set $(P(\theta,0), \partial_\tau P(\theta,0), Q(\theta,0), \partial_\tau Q(\theta,0), \lambda(0,0))=(p(\theta), \pi(\theta), q(\theta), \xi(\theta), \ell)$
must satisfy an integrability condition $\int_{S^1} (\pi \partial_\theta p +e^{2p} \xi \partial_\theta q )d\theta =0$. We let $\Pi$ denote the space of all such data sets. 

The extra field variable and the extra terms in the field equations  result in the dynamics of the Gowdy solutions being considerably more complicated than that of the polarized Gowdy. One can, however, still prove a model-SCC theorem for the general Gowdy $T^3$ solutions.\footnote{For the polarized  Gowdy solutions, one can prove a model-SCC theorem for all allowed Gowdy topologies; for the general Gowdy solutions, such has been proven only for the $T^3$ case.}\cite{RingSCC09}
\begin{theorem} [Model-SCC Theorem for $T^3$ Gowdy Spacetimes]
\label{SCCGowdy}
There exists a subset $\tilde \Pi$ which is open with respect to the $C^1 \times C^0$ topology in $\Pi$, and dense in the this space with respect to the $C^\infty$ topology, such that the maximal globally hyperbolic spacetime development of any data set in $\tilde \Pi$ is $C^2$ inextendible.
\end{theorem}

We note that the statement of the model-SCC theorem for $T^3$ Gowdy spacetimes is very similar to that for polarized Gowdy solutions. As well, the source  of the inextendibilty of both sorts of spacetimes is very similar: In both cases, one proves that the solutions are geodesically complete in the expanding direction ($\tau \rightarrow -\infty$), and that they generically have curvature blowups in the direction toward the shrinking (singular) direction ($\tau \rightarrow \infty$).

The proof of Theorem \ref{SCCGowdy} is, however, considerably more difficult than that of Theorem \ref{SCCPoldGowdy}. In a phenomenological sense, the source of this difficulty can be seen in a characteristic behavior which was first observed in Gowdy solutions in the numerical simulations of these spacetimes carried out by Berger and Moncrief in \cite{BM93} in the early 1990s. While their simulations did indicate the presence of AVTD behavior in the  $T^3$ Gowdy spacetimes, they also found  that very pronounced \emph{spikes} in the graphs of the metric fields often develop in the evolving spacetimes in a way which could in principle interfere with the asymptotic behavior expected in a spacetime characterized by AVTD behavior, should the spikes become very prevalent. A key feature of  Ringstr\"om's beautiful work \cite{Ring04, Ring05, RingSCC09} on the Gowdy spacetimes, work which culminated in a proof of Theorem \ref{SCCGowdy},  is the very careful treatment of these spikes in their many different forms. 

Indeed, an essential part of what  distinguishes $\tilde \Pi$ from $\Pi$ is the control of the formation of spikes. In solutions which evolve from data in $\tilde \Pi$, only a finite number of spikes develop. As a result, AVTD-type asymptotic behavior is observed along generic timelike paths approaching the singularity. Furthermore, along such paths, the asymptotic velocity 
\begin{equation}
\label{AsymptVeloc} 
V(\theta) := \lim_{\tau \rightarrow \infty} [\partial_\tau P^2 (\theta, \tau) + e^{2P} \partial_\tau Q^2 (\theta, \tau)]^{\frac{1}{2}}
\end{equation}
is well-defined \cite{Ring06a}. This quantity, which generalizes the asymptotic data function $v(\theta)$ used in working with polarized Gowdy spacetimes, determines whether or not the curvature is bounded along a timelike path which approaches a specified value of the coordinate $\theta$: The curvature blows up along such a path so long as the value of $V$ at that point is not one. This condition $V(\theta) \neq 1$ is found to hold generically, from which it follows that there can be no extensions across a Cauchy horizon (in the $\tau \rightarrow \infty$ direction) in the spacetimes corresponding to $\tilde \Pi$ data. Ringstr\"om proceeds to show that $\tilde \Pi$ is an open and dense subset of $\Pi$. This result, together with his verification that the $\tilde \Pi$ solutions are geodesically complete in the $\tau \rightarrow - \infty$ direction, proves Theorem \ref{SCCGowdy}.

\section{Evidence for AVTD Behavior in More General Families of Spacetimes}
\label{AVTD}

The $T^3$ Gowdy spacetimes constitute the least restrictive family of solutions of the vacuum Einstein's equations for which a model SCC theorem has been proven. Such theorems have been proven for more restrictive families of solutions, such as those which are spatially homogeneous (and therefore have a three-dimensional isometry group) \cite{Ren94, Chr-Ren95}. They have also been proven for a number of families of spacetimes satisfying various Einstein-matter equations, including polarized Gowdy solutions of the Einstein-Maxwell equations \cite{Nun-Ren09}, as well as  solutions of the Einstein-Vlasov equations with $T^2$-symmetry \cite{Daf-Ren06}, spherical symmetry, or hyperbolic symmetry \cite{Daf-Ren07}. Our main interest here is on what we know and what we conjecture for vacuum solutions with less restrictive conditions than the $T^3$ Gowdy solutions. 

In proving model SCC theorems for both the polarized and the general ($T^3$) Gowdy spacetimes, AVTD behavior plays an important role. Hence, in looking for more general families  of spacetimes for which such theorems may hold, it is useful to determine if there are such families for which AVTD behavior is known to be present. This is the case for three families of vacuum solutions: the polarized (and half-polarized) $T^2$-symmetric spacetimes, the polarized (and half-polarized) $U(1)$-symmetric spacetimes, plus general spacetimes in $(10+1)$-dimensions.  

It has not been proven for any of these families of solutions that AVTD behavior is to be found in every member of the family, or in some open and dense subset of the full family. However, for the polarized $T^2$-symmetric solutions  as well as for the polarized $U(1)$-symmetric solutions, there is strong evidence for AVTD behavior based on numerical simulations \cite{Lim}, \cite{BM98}. For all three families, it has been proven using Fuchsian methods that there are at least some solutions (a collection parametrized by the free choice of certain functions, in each case) with AVTD behavior. 

To illustrate how the Fuchsian approach works, we focus on the application of these techniques to the polarized $T^2$-symmetric solutions. Like the Gowdy solutions, the $T^2$-symmetric solutions are characterized by a  2-torus isometry group acting spatially. For the latter family, however, the twist constants do not vanish. As a consequence, the metrics necessarily take a more complicated form, which we write as follows:
 \begin{equation}
 \label{PolT2Metric}
    g = e^{2(\eta -U)} ( -\alpha dt^2 + d\theta^2 ) 
    + e^{2U} dx^2 
    + e^{-2U} t^2 ( dy + G d\theta )^2.
  \end{equation}
We note that if $\alpha =1$ and if $G$ vanishes in \eqref{PolT2Metric}, these metrics reduce to polarized Gowdy metrics. It is convenient for the discussion of the Fuchsian analysis that we use a slightly different form of metric parametrization here than that used above in the discussion of the Gowdy solutions; in particular, we replace $\tau$ by the time coordinate $t:= e^{-\tau}$, so that the singularity occurs at $t=0$, and we also make small changes in the choice of the metric coefficients (replacing $P$ and $\lambda$ by the closely related  $U$ and $\eta$).

The vacuum Einstein field equations for the polarized $T^2$-symmetric spacetimes take the form
\begin{align}
\label{T2U}
\partial_{tt}U +\frac{1}{t} \partial_tU &= \alpha \partial_{\theta \theta}U +\frac{1}{2}\partial_\theta \alpha \partial_\theta U +\frac{1}{2\alpha} \partial_t \alpha \partial_t U,\\
\label{T2etat}
\partial_t \eta &=t \partial_t U^2 + t \alpha \partial_\theta U^2 +\frac{e^{2\eta}}{4t^3} \alpha K^2,\\
\label{T2etax} 
\partial_\theta \eta &=2 t \partial_t U \partial_\theta U - \frac{\partial_\theta \alpha}{2 \alpha},\\
\label{T2alpha}
\partial_t \alpha &= - \frac{e^{2 \eta}}{t^3} \alpha^2 K^2,\\
\label{T2G}
\partial_t G &= e^{2 \eta} \sqrt{\alpha}K t^{-3},
\end{align}
where $K$ designates the non-vanishing twist constant which distinguishes these spacetimes from the Gowdy solutions\footnote{Without loss of generality in studying these spacetimes, one may set one of the twist constants to zero, labeleing the remaining one as $K$.}. Comparing the features of this system of equations with those of the Gowdy equations \eqref{GPeqn}-\eqref{Glambdatau} above (and the corresponding polarized Gowdy equations), we notice a key difference: While the Gowdy equations for $\lambda$ are semi-decoupled from those for $P$ and $Q$, here the system is fully coupled (apart from the equation \eqref{T2G} for $G$). In a rough sense, this coupling is responsible for making  the analysis of the polarized $T^2$-symmetric solutions more difficult (and consequently more interesting) than that of the polarized Gowdy solutions.

As noted above, numerical simulations strongly indicate that generic polarized $T^2$-symmetric solutions are AVTD in a neighborhood of their singularities at $t \rightarrow 0$. These simulations also suggest that the development of spikes, which complicate the dynamics of the general Gowdy solutions, plays at most a very minor role in the dynamics of polarized $T^2$-symmetric spacetimes. While no theorem concerning the generic presence of AVTD behavior in these spacetimes has yet been proven, Fuchsian methods have been used to show  that there are polarized $T^2$-symmetric solutions with AVTD behavior. Roughly speaking, the way this works is as follows.

Writing out the polarized $T^2$-symmetric VTD equations\footnote{These equations are clearly obtained by dropping the terms with spatial derivatives in the system \eqref{T2U}-\eqref{T2G}, \emph{except} for in the constraint equation \eqref{T2etax}, which contains no temporal derivatives.} as
\begin{align}
\label{T2UV}
\partial_{tt}U +\frac{1}{t} \partial_tU &=  \frac{1}{2\alpha} \partial_t \alpha \partial_t U,\\
\label{T2etatV}
\partial_t \eta &=t \partial_t U^2 +  +\frac{e^{2\eta}}{4t^3} \alpha K^2,\\
\label{T2etaxV} 
\partial_\theta \eta &=2 t \partial_t U \partial_\theta U - \frac{\partial_\theta \alpha}{2 \alpha},\\
\label{T2alphaV}
\partial_t \alpha &= - \frac{e^{2 \eta}}{t^3} \alpha^2 K^2,\\
\label{T2GV}
\partial_t G &= e^{2 \eta} \sqrt{\alpha} K t^{-3}, 
\end{align}
we first verify that for a general collection of functions $k(\theta)$, $U_*(\theta)$, $\eta_*(\theta)$, $\alpha_*$ and $G_*(\theta)$ (which we call collectively the ``asymptotic data"), the following are asymptotically solutions of this VTD system:
\begin{align}
  \label{VTDSolns}
  \hat U(\theta, t)&=\frac 12(1-k(\theta))\log t+U_{*}(\theta),\\
  \hat \eta(\theta, t)&=\frac 14(1-k(\theta))^2\log t+\eta_*(\theta),\\
  \hat \alpha(\theta, t)&=\alpha_*(\theta),\\
  \label{VTDSolnss}
  \hat G(\theta,t)&=G_*(\theta).
\end{align}
Next, we express the unknown metric coefficients as sums of these (function-parametrized) VTD solutions plus remainder-field terms $\tilde U(\theta,t), \tilde \eta(\theta,t), \tilde \alpha(\theta, t)$, and $\tilde G(\theta,t)$,
\begin{align}
\label{AVTDExpan}
U(\theta, t)&=\hat U (\theta, t) + \tilde U(\theta, t),\\
 \eta(\theta, t)&=\hat \eta(\theta, t) + \tilde \eta(\theta,t),\\ 
\alpha(\theta, t)&=\hat \alpha (\theta,t) + \tilde \alpha (\theta, t),\\
G(\theta,t)&=\hat G(\theta, t)+\tilde G(\theta, t), 
\end{align}
and we substitute these expressions into the polarized $T^2$-symmetric Einstein vacuum equations \eqref{T2U}-\eqref{T2G}. We thus obtain a ($k(\theta)$, $U_*(\theta)$, $\eta_*(\theta)$, $\alpha_*(\theta)$ $G_*(\theta)$)-parametrized PDE system for the remainder-field terms. 

The idea is to show that for each suitable choice of the asymptotic data, there exists (for $t$ sufficiently close to zero) a unique solution to the remainder-field PDE system, and to show moreover that all of the remainder fields $\tilde U(\theta,t), \tilde \eta(\theta,t), \tilde \alpha(\theta, t)$, and $\tilde G(\theta,t)$ included in this unique solution approach zero as $t$ approaches zero. If one can do this, it follows that the polarized $T^2$-symmetric spacetime composed from the specified asymptotic data together with the resulting remainder-field solution is AVTD. 

If the asymptotic data functions are all real analytic, it is relatively straightforward  to determine conditions on this data which are sufficient for the existence of a remainder-field solution with the desired decay. So long as one can write the remainder-field PDE system collectively in the form 
\begin{equation} 
\label{AnalytFuchs}
t\partial_t \Phi + M(\theta) \Phi = t^\epsilon F(\theta, t, \Phi, \partial_\theta \Phi),
\end{equation}
where the vector field $\Phi$ includes as its components all of the remainder fields and their $\theta$ derivatives, where the matrix $M$ (whose explicit form depends on the asymptotic data) must satisfy certain positivity conditions, where $\epsilon$ is a positive constant, and where the function F (also depending on the asymptotic  data) is continuous in $t$, is analytic in all of its other arguments, and extends continuously to $t=0$, then indeed a unique solution satisfying the desired properties exists. As shown in \cite{IK99}, so long as $\alpha_*(\theta)$ is positive, so long as $k(\theta)$ satisfies certain inequalities, and so long as the asymptotic data collectively satisfy an integrability condition (derived from the constraint equation \eqref{T2etax}), then the remainder field equations can be written in this form, with $M$ and $F$ satisfying the conditions listed above. Thus one verifies that there exists a parametrized set of real analytic polarized  $T^2$-symmetric solutions which exhibit AVTD behavior. 

Although Fuchsian techniques were originally developed to work with real analytic solutions of PDE systems with real analytic coefficients (see, e.g., \cite{Kich}), they have since been adapted (by Rendall in \cite{Ren00}, and by Ames, Beyer, LeFloch, and the author in  \cite{ABIL}) to apply to  PDE systems and solutions of those systems with much less assumed regularity. These adaptations (to date) require one to work with a more restricted class of PDE systems, such as those which are quasilinear and symmetric hyperbolic and take the form 
\begin{equation}
\label{QlinSymHyp}
S(\theta, t, \Psi)t \partial_t \Psi + T(\theta, t, \Psi) t \partial_\theta\Psi +N(\theta, t, \Psi) \Psi =E(\theta, t, \Psi),
\end{equation}
where $\Psi(\theta, t)$ is a vector-valued function representing the collection of fields and their first order derivatives.  As discussed in \cite{ABIL}, so long as a number of technical conditions are satisfied by the matrix functions $S, T, N$ and $E$,  both in general and for certain choices of $\Psi$ as ``asymptotic data" $\hat \Psi(\theta, t)$, then it follows that for those choices of the asymptotic data, there exist unique solutions $\Psi=\hat \Psi +\tilde \Psi$ of \eqref{QlinSymHyp} with the remainder terms $\tilde \Psi$ decaying to zero as $t \rightarrow 0$. 

Both for smooth PDE coefficients and asymptotic data, and for less regular choices of the coefficients and the asymptotic data (as specified by certain choices of weighted Sobolev spaces\footnote{The weighting pertains to the decay of functions as they approach $t=0$.}), these adapted Fuchsian techniques have been used to find parametrized classes of polarized $T^2$-symmetric solutions (of the stated regularity) with AVTD behavior near the $t=0$ singularity \cite{ABIL}. In doing this, one chooses $\hat \Psi(\theta, t)$ to correspond to the choices \eqref{VTDSolns}-\eqref{VTDSolnss} of $\hat U(\theta,t), \hat \eta(\theta,t), \hat \alpha(\theta, t)$, and $\hat G(\theta,t)$ discussed above, which asymptotically approach solutions of the VTD equations \eqref{T2UV}-\eqref{T2GV}. 

As is the case for the real analytic solutions, it is not yet known if these solutions with AVTD behavior represent anything more than a set of measure zero among all polarized $T^2$-symmetric solutions. Numerical simulations suggest that indeed AVTD behavior may be prevalent, if not generic \cite{Lim},  but nothing of this nature has been proven.

Fuchsian methods have been used to prove that other families of solutions of Einstein's equations include at least some solutions with AVTD behavior. The earliest such results pertained to the $T^3$ Gowdy solutions: Kichenassamy and Rendall used Fuchsian techniques to prove the existence of real analytic Gowdy solutions with AVTD behavior in \cite{KR98}, and Rendall did the same for smooth Gowdy solutions in \cite{Ren00}. This work of course presaged Ringstr\"om's proof that generic $T^3$ Gowdy solutions exhibit AVTD behavior. For the $T^2$-symmetric spacetimes, Fuchsian methods have been used to show not only that there are polarized solutions with AVTD behavior, but that there are``half-polarized" solutions with this behavior as well \cite{CI07}. Half-polarized solutions allow the presence of  a non vanishing $Q(\theta,t) dxdy$ term in the expression for the metric \eqref{PolT2Metric}; however the dynamics of this term is strongly restricted, with one non vanishing function in the asymptotic data controlling it, as opposed to the two functions ($k$ and $U_*$) in the asymptotic data which control $U$. 

The Gowdy spacetimes and the $T^2$-symmetric solutions are all characterized by the very restrictive assumption that they each admit  a spatially-acting two-dimensional isometry group. Loosening this restriction to the admission of a spatially-acting isometry group of only one dimension, one finds that Fuchsian methods can  indeed be used to prove that among these $U(1)$-symmetric spacetimes, there are some which show AVTD behavior.  As for the $T^2$-symmetric spacetimes, AVTD behavior has been shown to exist only in $U(1)$-symmetric solutions which are either polarized or half-polarized\cite{IM02, CBIM04}.  We note as well that  all $U(1)$-symmetric spacetimes shown thus far to exhibit AVTD behavior are real analytic. 

It is expected that using Fuchsian methods we will be able to show that there are smooth polarized $U(1)$-symmetric solutions with AVTD behavior. Based on evidence from numerical simulations \cite{BM00}, it is \emph{not}, however, expected that  $U(1)$-symmetric solutions without any polarization restriction  will be found which exhibit AVTD behavior. The same is true for $T^2$-symmetric solutions \cite{BIW01}. 

There has been significant speculation that for families of solutions more general than those discussed thus far, while AVTD behavior may not be found, one may find ``Mixmaster" behavior instead. Roughly speaking, a solution shows Mixmaster behavior near its singularity if observers approaching  the singularity do not each see his/her own Kasner-like behavior\footnote{We recall that the VTD equations involve the dropping of all  spatial-derivative terms in the Einstein evolution equations; hence the metric evolution seen by each observer is the same as the evolution of a spacetime with a spatially-acting $T^3$-isometry group, which corresponds to the Kasner spacetime.}, but rather each sees his/her own Bianchi type IX solution\footnote{These are the solutions with $SU(2)$ acting transitively on space-like slices.}. The Mixmaster evolution is characterized by an infinite succession of episodic Kasner-type evolutions, each of which is ultimately disrupted by a short-lived ``bounce", followed by a transition to the next Kasner episode. 

The prediction that Mixmaster behavior is likely to be seen generically in spacetime solutions less restricted than those which exhibit AVTD behavior is based partly on numerical simulations (as cited above \cite{BIW01, BM00}), partly on the pioneering work of Belinskii, Khalatnikov and Lifschitz (BKL)\cite{BKL70}, and partly on more recent speculative studies \cite{Dd08}. On the other hand, others have  argued that the prevalence of spikes in the evolution of these spacetimes with more intricate dynamics strongly indicate that the conjecture of generic Mixmaster behavior is very unlikely to hold. This issue remains to be settled. 

We do note that, should it be shown that Mixmaster behavior characterizes the behavior of generic solutions near their singularities, in light of the fairly good understanding we have of the evolution of the curvature in Bianchi Type IX solutions, such a result could be a very useful tool for the study of Strong Cosmic Censorship. 

We close this section by noting that, if instead of vacuum solutions one considers spacetimes satisfying the Einstein equations with certain stiff fluids or scalar fields coupled in, then AVTD behavior \emph{is} found. This was shown using Fuchsian methods (applied to real analytic solutions) by Andersson and Rendall in \cite{AR01}. More recently, Rodnianski and Speck have shown \cite{RS} that the presence of AVTD behavior is in fact \emph{stable} among these Einstein-scalar field or Einstein-stiff fluid solutions. We also note the work \cite{DHRW} which uses Fuchsian techniques to prove that there are vacuum solutions of dimension 11 or higher which show AVTD behavior.

\section{Blue Shift Effects, Weak Null Singularities, and the Nature of SCC}
\label{Blue}

While some of the original motivation for believing in the validity of the Strong Cosmic Censorship conjecture  came from the conviction that a respectable theory of the gravitational field should not (generically) allow for physical spacetimes in which one's ability to predict the future from knowledge about the past breaks down, Penrose also based his view that SCC should hold on his assessment of the ``blue shift effect" on Cauchy horizons in black hole interiors. Reissner-Nordstom as well as Kerr black hole interiors contain Cauchy horizons; if the existence of these structures were found to be stable under generic perturbations, then SCC would be refuted. However, Penrose reasoned that any small perturbation of an astrophysical system evolving towards a Reissner-Nordstrom or a Kerr spacetimes would ``fall" into the developing black hole, and in doing so would become strongly blue-shifted and consequently would become powerful enough to destroy the Cauchy horizon. This blue shift effect would  therefore ``save" Strong Cosmic Censorship. 

Early, somewhat heuristic, explorations  of this blue shift effect (by Hiscock, Israel, Poisson, Ori, and others) have suggested a surprising scenario: that perturbations of charged black hole solutions would contain null surfaces across which \emph{continuous extensions of the metric} could be made, but \emph{continuous extensions of the curvature} could not be made. That is, the Reissner-Nordstrom Cauchy horizons, according to this scenario, are stable in a certain $C^0$ sense, but not in a $C^2$ sense. 

Remarkably, a wide range of subsequent studies very strongly support this scenario. In the first of these works (done twelve years ago), Dafermos \cite{D03, D05} has shown that for any asymptotically Euclidean spherically symmetric initial data set with non vanishing charge whose Einstein-Maxwell development forms a black hole, the maximal globally hyperbolic development does admit a non-empty null surface across which a continuous extension of the metric can be carried out. Further (with certain technical assumptions), he proves \cite{D05} that these $C^0$-type Cauchy horizons are generically singular, with the curvature and the Hawking mass blowing up; moreover, the metric extensions do not admit locally square integrable Christoffel symbols. These properties have led to these null surfaces being labeled  \emph{weak null singularities}. 

Considerations of this work raise three important questions: Do weak null singularities exist in spacetimes which are not spherically symmetric? If so, might they characterize generic perturbations of Reissner-Nordstrom and Kerr (and Kerr-Newman) black holes and their interiors? If weak null singularities are stable, does this constitute a proof that Strong Cosmic Censorship is false?

The first question is answered by very recent work by Luk \cite{Luk}, which proves that the maximal developments of certain sets of characteristic initial data (which are \emph{not} required to admit any isometries) always contain weak null singularities. While the second question has not been fully resolved in general, Dafermos and Luk claim that they can show that \emph{if} the stability of the exterior structure of the Kerr solutions can be proven\footnote{There is, of course, a very large amount of mathematical effort currently being directed toward proving the stability of Kerr solutions.}, then the stability of the internal Cauchy horizon, as a weak null singularity, would follow as a corollary. 

Should we now conclude that if Kerr is proven to be stable (in the exterior sense), then the Strong Cosmic Censorship conjecture is false? This becomes a matter of interpretation. Since its inception, Strong Cosmic Censorship has been an imprecise and malleable conjecture. To formulate it explicitly, one needs to answer all of the following questions: What are ``generic solutions"? Is the primary issue whether or not the curvature is bounded in the neighborhood of the singularity? Or is the issue whether or not spacetime metric extensions can be carried out? If the stability of extensions is the key, then does it matter whether the curvature as well as the metric can be extended?  Does it matter if the singularity which forms is spacelike or null?

Resolving all of these questions is important if one wants a single statement of the Strong Cosmic Censorship conjecture, to confirm or refute. On the other hand, it may be more useful to consider several different versions of the conjecture, and ultimately determine that there are some reasonable forms of SCC that are true, and others which are not. 

The primary purpose of the Strong Cosmic Censorship conjecture has always been to stimulate interesting questions and studies in mathematical relativity. For this purpose, it has certainly been successful.

\section*{Acknowledgements}
This work was partially supported by NSF grant  PHY-1306441 at the University of Oregon. I thank the Simons Center for hospitality during the course of the writing of this review, and I thank Beverly Berger, Mihalis Dafermos, Ellery Ames and Florian Beyer for useful conversations. I also thank the referee for useful comments.

\addcontentsline{toc}{section}{References}
\bibliography{bibliography}

\end{document}